\documentstyle[aps,prl,twocolumn,floats,psfig,epsf]{revtex}
\begin{document}
\draft
\title{{\rm PHYSICS }\hfill {\sl Version of \today}\\~~\\
Random Noise and Pole-Dynamics in Unstable Front Propagation(new
version)}
\author {Oleg Kupervasser, Zeev Olami, Barak Galanti  and
Itamar  Procaccia}
\address{Department of~~Chemical Physics,\\
 The Weizmann Institute of Science,
Rehovot 76100, Israel}
\maketitle
\begin{abstract}
The problem of flame propagation is studied as an example of unstable fronts that
wrinkle on many scales. The analytic tool of pole expansion
in the complex plane is employed to address the interaction of the unstable growth
process with random initial conditions and
perturbations. We argue that the effect of random noise is immense and that it
can never be neglected in sufficiently large systems. We present simulations
that lead to scaling laws
for the velocity and acceleration of the front as a function of the system
size and the level of noise, and analytic arguments that explain these results
in terms of the noisy pole dynamics.
\end{abstract}
\pacs{PACS numbers 47.27.Gs, 47.27.Jv, 05.40.+j}
\section{Introduction}
In our last paper about flame front propagation we left two open problems.
The first one is how to explain the existence of small dependence parameters
of problem on the noise. The second problem is how to calculate numerically
such values as excess number of poles in system , number of poles that appear
in the system in unit of time, life time of pole.To solve these problem
we write this paper.

\section{Equations of Motion and Pole-decomposition in the Channel Geometry}
It is known that planar flames freely propagating
through initially motionless homogeneous combustible mixtures are
intrinsically unstable.
It was reported that such flames develop characteristic structures which
include cusps,
and that under usual experimental conditions the flame front accelerates as
time goes
on. A model in $1+1$ dimensions that pertains to the propagation of flame
fronts in channels of width
$\tilde L$ was proposed in \cite{77Siv}. It is written in terms of position
$h(x,t)$ of the
flame front above the $x$-axis. After appropriate rescalings it takes
the form:
\begin{equation}
{\partial h(x,t) \over \partial t}=
{1\over 2}\left[{\partial h(x,t) \over \partial x }\right]^2
 +\nu{\partial^2 h(x,t)\over \partial x^2}+ I\{h(x,t)\}+1 \ . \label{Eqnondim}
\end{equation}
The domain is  $0 <x< \tilde L$, $\nu$ is a parameter and we use periodic boundary
conditions. The functional $I[h(x,t)]$
is the Hilbert transform of derivative
which is conveniently defined in terms of the spatial
Fourier transform
\begin{eqnarray}
&&h(x,t)= \int_{-\infty}^{\infty} e^{i k x}\hat h(k,t) dk \label{Four}\\
&& I[h(k,t)] = |k| \hat h(k,t) \label{hil}
\end{eqnarray}
For the purpose of introducing the pole-decomposition it is convenient to
rescale the domain to $0<\theta <2\pi$. Performing this rescaling and
denoting the
resulting quantities with the same notation we have
\begin{eqnarray}
&&{\partial h(\theta,t) \over \partial t}=
{1\over 2L^2}\left[{\partial h(\theta,t) \over \partial \theta }\right]^2
 +{\nu\over L^2}{\partial^2 h(\theta,t)\over \partial\theta^2}\nonumber \\&&+
{1\over L}I\{h(\theta,t)\}+1 \ .
\label{Eqdim}
\end{eqnarray}
In this equation $L=\tilde L/2\pi$.
Next we change variables to $u(\theta,t)\equiv {\partial
h(\theta,t)/\partial\theta}$.
 We find
\begin{equation}
{\partial u(\theta,t) \over \partial t}=
{u(\theta,t)\over L^2}{\partial u(\theta,t) \over \partial \theta }
 +{\nu\over L^2}{\partial^2 u(\theta,t)\over \partial \theta^2}+ {1\over
L}I\{u(\theta,t)\} \ . \label{eqfinal}
\end{equation}
It is well known that the flat front solution of this equation is linearly
unstable.
The linear spectrum in $k$-representation is
\begin{equation}
\omega_k=|k|/L-\nu k^2/L^2 \ . \label{spec}
\end{equation}
There exists a
typical scale $k_{max}$ which is the last unstable mode
\begin{equation}
k_{max} = {L\over \nu} \ . \label{kmax}
\end{equation}
Nonlinear effects stabilize a new steady-state which is discussed next.

The outstanding feature of the solutions of this equation is the appearance
of cusp-like structures in the developing fronts. Therefore a representation in
terms of Fourier modes is very inefficient. Rather, it appears very
worthwhile to represent such solutions in terms of sums of functions of poles
in the complex plane. It will be shown below that the position of the cusp along the
front is determined by the real coordinate of the pole, whereas the height of the
cusp is in correspondence with the imaginary coordinate. Moreover, it will
be seen that the dynamics of the developing front can be usefully described in
terms of the dynamics of the poles. Following \cite{82LC,85TFH,90Jou,96KOP} we
expand the solutions $u(\theta,t)$ in functions that depend on
$N$ poles whose position $z_j(t)\equiv x_j(t)+iy_j(t)$ in the complex plane
is time dependent:
\begin{eqnarray}
&&u(\theta,t)=\nu\sum_{j=1}^{N}\cot \left[{\theta-z_j(t) \over 2}\right]
   + c.c.\nonumber \\
&&=\nu\sum_{j=1}^{N}{2\sin [\theta-x_j(t)]\over
\cosh [y_j(t)]-\cos [\theta-x_j(t)]}\ , \label{upoles}
\end{eqnarray}
\begin{equation}
h(\theta,t)=2\nu\sum_{j=1}^{N}{\ln \Big[\cosh (y_j(t))-\cos (\theta-x_j(t))
\Big]}+C(t) \ . \label{rpoles}
\end{equation}
In (\ref{rpoles}) $C(t)$ is a function of time. The function (\ref{rpoles})
is a superposition of quasi-cusps (i.e. cusps that are rounded at the tip). The
real part of the pole position (i.e. $x_j$) is the coordinate (in the domain
$[0,2\pi]$) of the maximum
of the quasi-cusp, and the imaginary part of the pole position (i.e $y_j$)
is related to
the depth of the quasi-cusp. As $y_j$ decreases the depth of the cusp
increases. As $y_j \to 0$ the depth diverges to infinity. Conversely, when $y_j\to
\infty$ the
depth decreases to zero.

The main advantage of this representation is that the propagation and
wrinkling of the
front can be described via the dynamics of the poles. Substituting
(\ref{upoles}) in (\ref{eqfinal}) we derive the following ordinary
differential equations for the positions of the poles:
\begin{equation}
- L^2{dz_{j}\over dt}=\Big[\nu\sum_{k=1
  ,k\neq j}^{2N }\cot \left({z_j-z_k\over 2}\right)
  +i{L\over 2 }sign [Im(z_j)]\Big].\label{eqz}
\end{equation}
We note that in (\ref{upoles}), due to the complex conjugation, we have $2N$ poles
which are arranged in pairs such that for $j<N$ $z_{j+N}=\bar z_j$. In the
second sum in (\ref{upoles}) each pair of poles contributed one term. In Eq.(\ref{eqz})
we again employ $2N$ poles since all of them interact. We can write the
pole dynamics in  terms of the real and imaginary parts $x_j$ and $y_j$.
Because of the
arrangement in pairs it is sufficient to write the equation for either
$y_j>0$ or for
$y_j<0$. We opt for the first. The equations for the positions of the poles read
\begin{eqnarray}
&&-L^2{dx_{j}\over dt}=\nu\sum_{k=1,k\neq j}^N
   \sin(x_j-x_k)\Bigg[
   [\cosh (y_j-y_k) \label{xj} \\ && -\cos (x_j-x_k)]^{-1}+[\cosh
(y_j+y_k)-\cos (x_j-x_k)]^{-1}\Bigg]  \nonumber\\
&& L^2{dy_{j}\over dt}=\nu\sum_{k=1,k\neq j}^{N }\Big({\sinh (y_j-y_k)\over
   \cosh (y_j-y_k)-\cos (x_j-x_k)}\nonumber \\ &&+
   {\sinh (y_j+y_k)\over \cosh (y_j+y_k)-\cos (x_j-x_k)}
   \Big)+\nu\coth (y_j)- L\label{yj}  .
\end{eqnarray}
We note that if the initial conditions of the differential equation
(\ref{eqfinal})
are expandable in a finite number of poles, these equations of motion
preserve this number
as a function of time. On the other hand, this may be an unstable situation
for the
partial differential equation, and noise
can change the number of poles. This issue will be examined at length in
Section
\ref{noise}. We turn now to a discussion of the steady state solution of
the equations
of the pole-dynamics.
\subsection{Qualitative properties of the stationary solution}

The steady-state solution of the flame front propagating in channels of
width $2\pi$
was presented in Ref.\cite{85TFH}. Using these results we can immediately
translate
the discussion to a channel of width $L$. The main results are summarized
as follows:
\begin{enumerate}
\item There is only one
stable stationary solution which is geometrically represented by a giant
cusp (or equivalently one finger) and
analytically by $N(L)$ poles which are aligned on one line parallel to the
imaginary
axis. The existence of this solution is made clearer with the following remarks.
\item There exists an attraction between the poles
along the real line. This is obvious from Eq.(\ref{xj}) in which the sign
of $dx_j/dt$ is
always determined by $\sin(x_j-x_k)$. The resulting dynamics merges all the $x$
positions of poles whose $y$-position remains finite.
\item The $y$
positions are distinct, and the poles are aligned above each others in positions
$y_{j-1}<y_j<y_{j+1}$ with the maximal being $y_{N(L)}$. This can be understood
from Eq.(\ref{yj}) in which the interaction is seen to be repulsive at
short ranges,
but changes sign at longer ranges.
\item If one adds
an additional pole to such
a solution, this pole (or another) will be pushed to infinity along the
imaginary axis.
If the system has less than $N(L)$ poles it is unstable to the addition of
poles,
and any noise will drive the system towards this unique state. The number
$N(L)$ is
\begin{equation}
N(L)= \Big[{1 \over 2}\left( {L \over \nu }+1\right) \Big]\ , \label{NofL}
\end{equation}
where $\Big[ \dots \Big]$ is the integer part. To see this consider
a system with $N$ poles and such
that all the values of $y_j$ satisfy the condition $0< y_j <y_{max}$.
Add now one additional pole whose coordinates are $z_a\equiv (x_a,y_a)$ with
$y_a\gg
y_{max}$. From the equation of motion for $y_a$, (\ref{yj}) we see that
the terms in the sum are all of the order of unity as is also the
$\cot(y_a)$ term. Thus
the equation of motion of $y_a$ is approximately
\begin{equation}
{dy_a \over dt}\approx \nu{2N+1 \over L^2}-{1\over L} \ . \label{ya}
\end{equation}
The fate of this pole depends on the number of other poles.
If $N$ is too large the pole will run to infinity, whereas if $N$ is  small
the pole
will be attracted towards the real axis. The condition for moving away to infinity
is that $N > N(L)$ where $N(L)$ is given by (\ref{NofL}). On the
other hand the $y$ coordinate of the poles cannot hit zero.
Zero is a repulsive line, and poles are pushed away from zero with infinite velocity.
 To see this consider
a pole whose $y_j$ approaches zero. For any finite $L$ the term
$\coth(y_j)$ grows
unboundedly whereas all the other terms in Eq.(\ref{yj}) remain  bounded.
\item  The height of the cusp is proportional to $L$. The distribution of
positions
of the poles along the line of constant $x$ was worked out in \cite{85TFH}.
\end{enumerate}
We will refer to the solution with all these properties as the
Thual-Frisch-Henon
(TFH)-cusp solution.

\section{Acceleration of the Flame Front, Pole Dynamics and Noise}
\label{noise}
A major motivation of this Section is the observation that in radial
geometry the
same equation of motion shows an acceleration of the flame
front. The aim of this section is to argue that this phenomenon is caused by the
noisy generation of new poles. Moreover, it is our contention that a great
deal can
be learned about the acceleration in radial geometry by considering the
effect of
noise in channel growth. In Ref. \cite{85TFH} it was shown that any initial
condition
which is represented in poles
goes to a unique stationary state which is the giant cusp which propagates with
a constant velocity $v=1/2$ up to small $1/L$ corrections. In light of our
discussion
of the last section we expect that any smooth enough initial condition will
go to
the same stationary state. Thus if there is no noise in the dynamics of a finite
channel, no acceleration of the flame front is possible. What happens if we
add noise to the system?

For concreteness we introduce an additive white-noise term $\eta(\theta,t)$to
the equation of motion  (\ref{eqfinal})  where
\begin{equation}
\eta(\theta,t) = \sum_k{\eta_k(t) \exp{(ik\theta)}}\ , \label{eta}
\end{equation}
and the Fourier amplitudes $\eta_k$ are correlated according to
\begin{equation}
<\eta_k(t)\eta^{*}_{k'}(t')>={f \over L}
\delta_{k,k'}\delta(t-t') \ . \label{corr}
\end{equation}
We will first examine the result of numerical simulations of noise-driven
dynamics,
and later return to the theoretical analysis.
\subsection{Noisy Simulations}
\begin{figure}
\epsfxsize=9.0truecm
\epsfbox{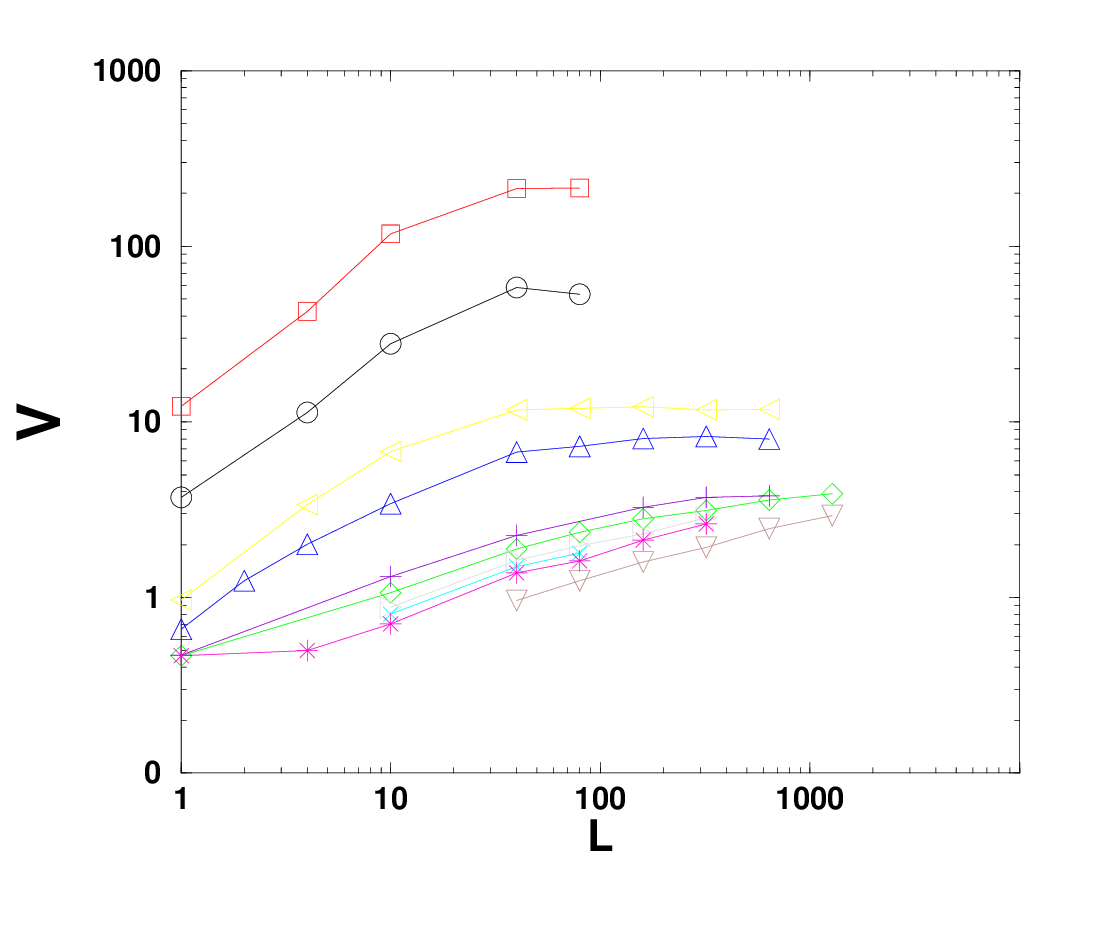}
\caption
{The dependence of the average velocity $v$ on the system size $L$
for $f^{0.5}=0, 2.7e-6, 2.7e-5, 2.7e-4, 2.7e-3, 2.7e-2, 2.7e-1,
0.5, 1.3, 2.7$.}
\label{file=Fig.2}
\end{figure}
\begin{figure}
\epsfxsize=9.0truecm
\epsfbox{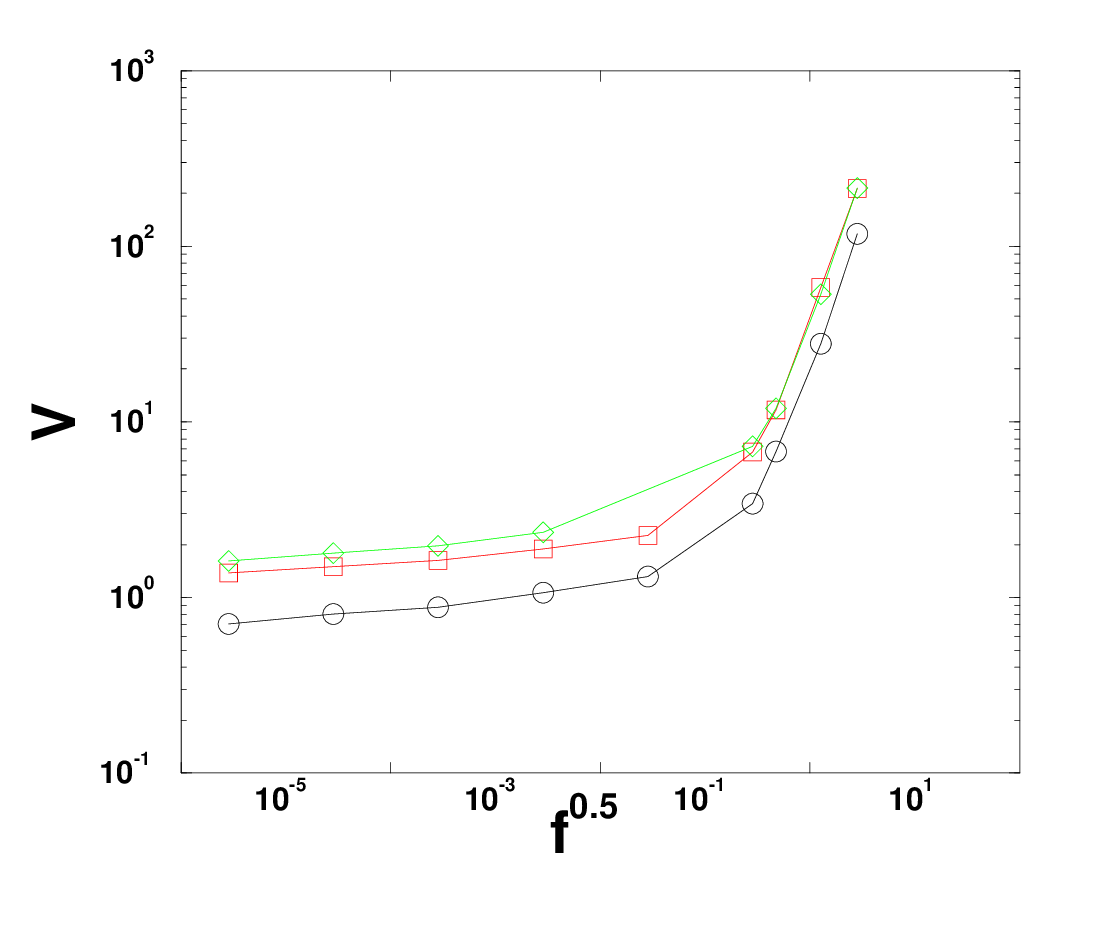}
\caption
{The dependence of the average velocity $v$ on the noise $f^{0.5}$ for L=10, 40, 80.}
\label{file=Fig.2}
\end{figure}

Previous numerical  investigations \cite{94FSF,90GS} did not introduce noise in
a controlled
fashion. We will argue later that some of the phenomena encountered in
these simulations
can be ascribed to the (uncontrolled) numerical noise. We performed numerical
simulations of Eq.(\ref{eqfinal} using a pseudo-spectral method. The time-stepping scheme
was chosen as Adams-Bashforth with 2nd order presicion in time. The additive white noise was
generated in Fourier-space by choosing
$\eta_k$
for every $k$ from a flat distribution
in the interval $[-\sqrt{2{f \over L}},\sqrt{2{f \over L}}]$.
We examined the average steady state
velocity of the front
as a function of $L$ for fixed $f$ and as a function of $f$ for fixed $L$.
We found the
interesting phenomena that are summarized here:
\begin{enumerate}
\item
In Fig.2 we can see two different regimes of  behavior the average velocity
$v$ as function of noise
$f^0.5$ for fixed system size L. For the noise $f$ smaller then same fixed value
$f_{cr}$
\begin{equation}
v\sim f^\xi \ . \label{vf}
\end{equation}
For these values of $f$ this dependence is very weak, and
$\xi\approx 0.02$.
 For
 large values of $f$ the dependence is much stronger
\item
In Fig.1 we can see growth of the average velocity $v$ as function of
the system size L. After some values of L we can see saturation of the
velocity.
For regime $f<f_{cr}$ the growth of the velocity can be written as
\begin{equation}
v\sim L^\mu , \quad \mu\approx 0.40\pm 0.05 \ . \label{scale1}
\end{equation}

\end{enumerate}
\subsection{Calculation  of Poles Number in the System}
The interesting problem that we would like to solve here it is to find
number of poles that exist in our system outside the giant cusp.
We can make it by next way: to calculate number of cusps (points of
minimum or inflexional points) and their position on the interval $
\theta:[0,2\pi]$ in every moment of time and to draw positions of cusp
like function of time, see Fig.3.
\begin{figure}
\epsfxsize=9.0truecm
\epsfbox{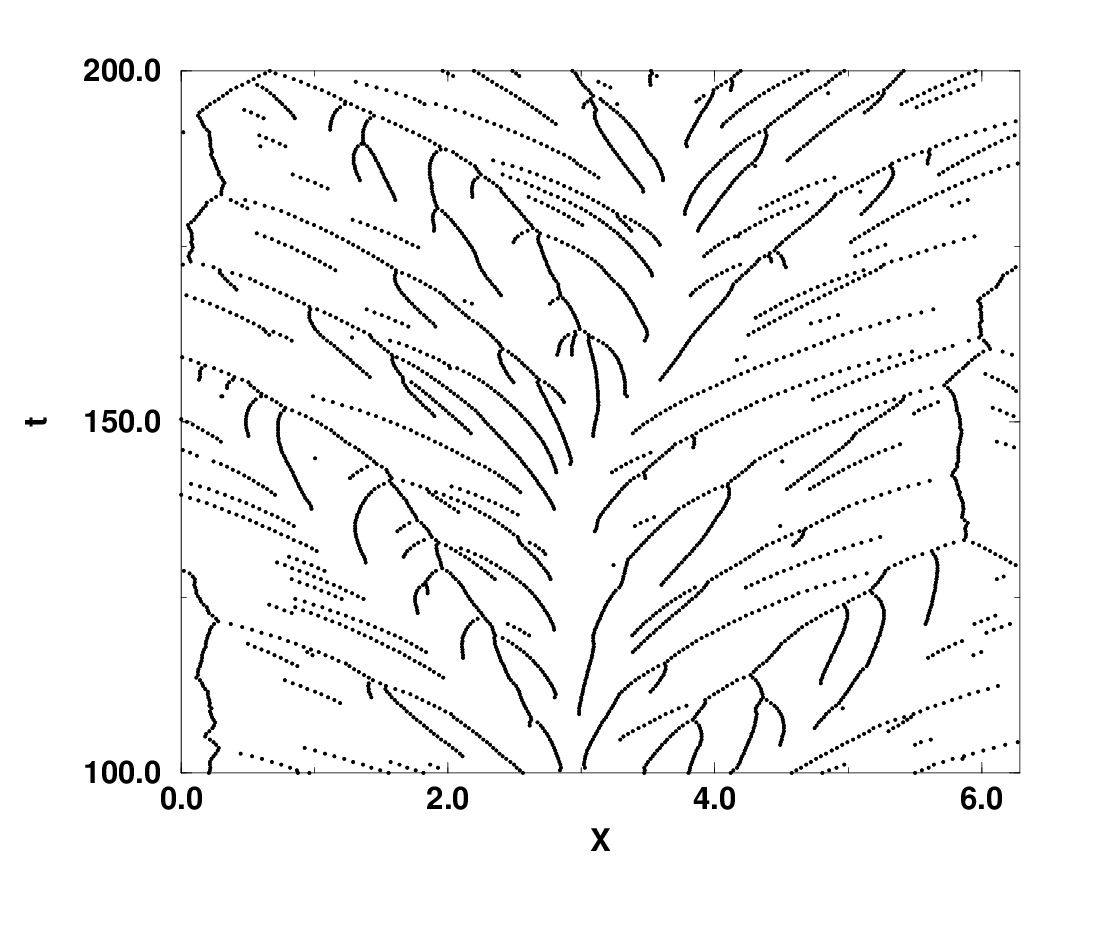}
\caption
{The dependence of the cusps positions  on time.$L=80$ $\nu=0.1$ $f=9e-6$}
\label{file=Fig.2}
\end{figure}

We assume that our system is almost
all time in "quasi-stable" state, i.e. every new cusp that appears in
the system includes only one pole.
By help pictures obtained by such way we can find
\begin{enumerate}
\item By calculation number of cusp in some moment of time and
by investigation of history of every cusp (except the giant cusp)
, i.e. how many initial cusps
take part in formation this cusp, after averaging with respect to
different moments of time we can find  mean number of poles that
exist  in our system outside the giant cusp.
Let us denote this number $\delta N$.
We can see four regimes that can be define with respect to dependence of
this number on noise $f$:

(i) Regime I: Such small noise that
no poles exist in our system outside the giant cusp.

(ii)
Regime II Strong dependence of poles number  $\delta N$  on noise $f$.

(iii) Regime III Saturation  poles number  $\delta N$  on noise $f$,
so we see very small dependence of this number on noise

\begin{equation}
\delta N \sim   f^{0.03}
\end{equation}
\begin{figure}
\epsfxsize=9.0truecm
\epsfbox{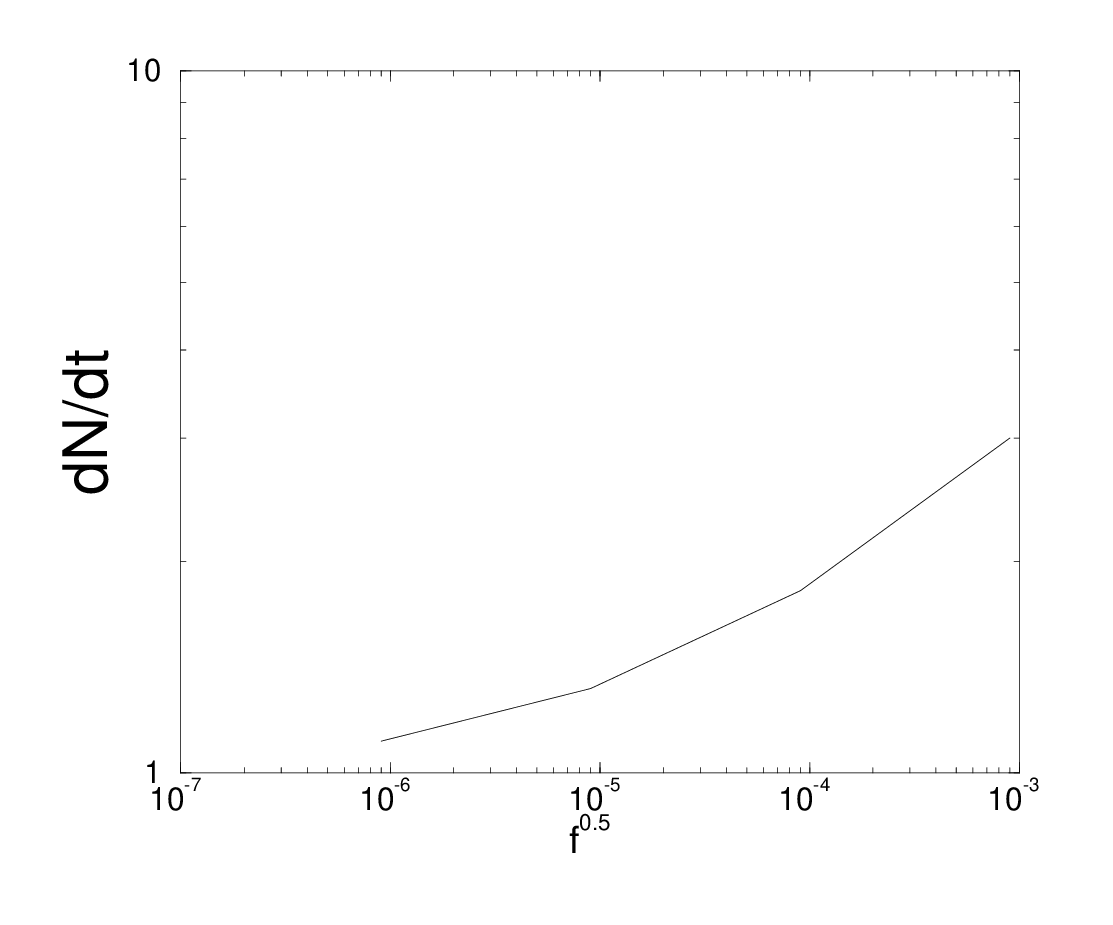}
\caption
{The dependence of poles number in time unit $dN/dt$ on the noise $f^{0.5}$.
$\nu=0.1$ $ L=80$}
\label{file=Fig.2}
\end{figure}
\begin{figure}
\epsfxsize=9.0truecm
\epsfbox{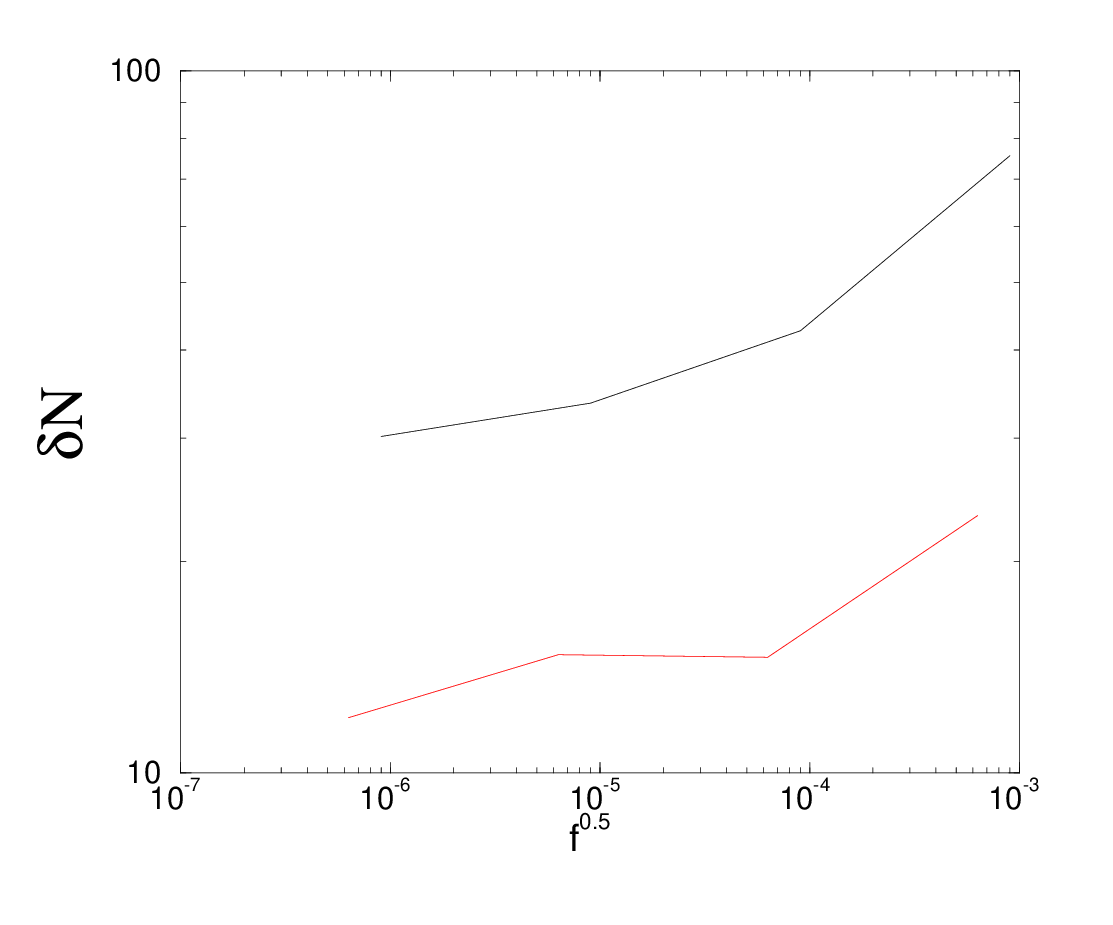}
\caption
{The dependence of excess poles number $\delta N$ on the noise $f^{0.5}$.
$\nu=0.1$ $ L=40,80$.}
\label{file=Fig.2}
\end{figure}
\begin{figure}
\epsfxsize=9.0truecm
\epsfbox{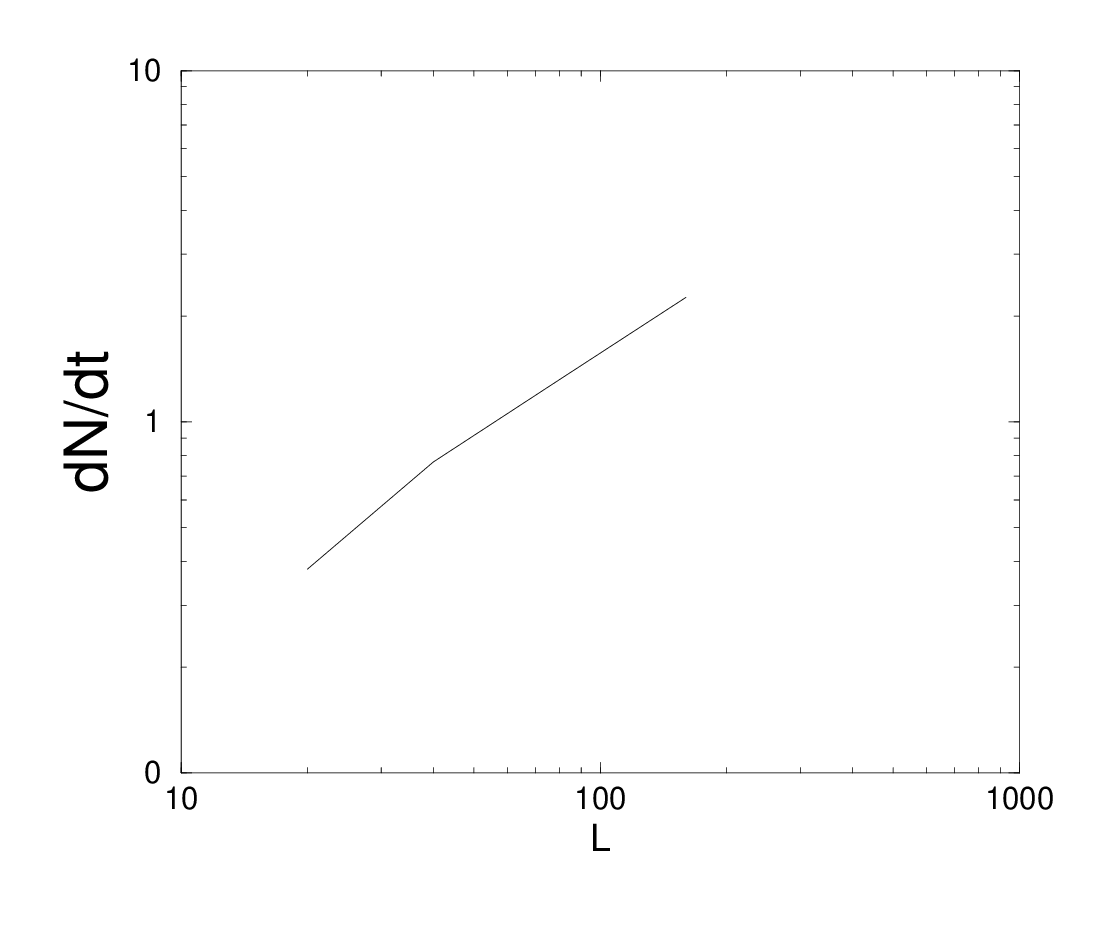}
\caption
{The dependence of poles number in time unit $dN/dt$ on the system size L.
$\nu=0.1$ $ f^{0.5}=9e-6$.}
\label{file=Fig.2}
\end{figure}
\begin{figure}
\epsfxsize=9.0truecm
\epsfbox{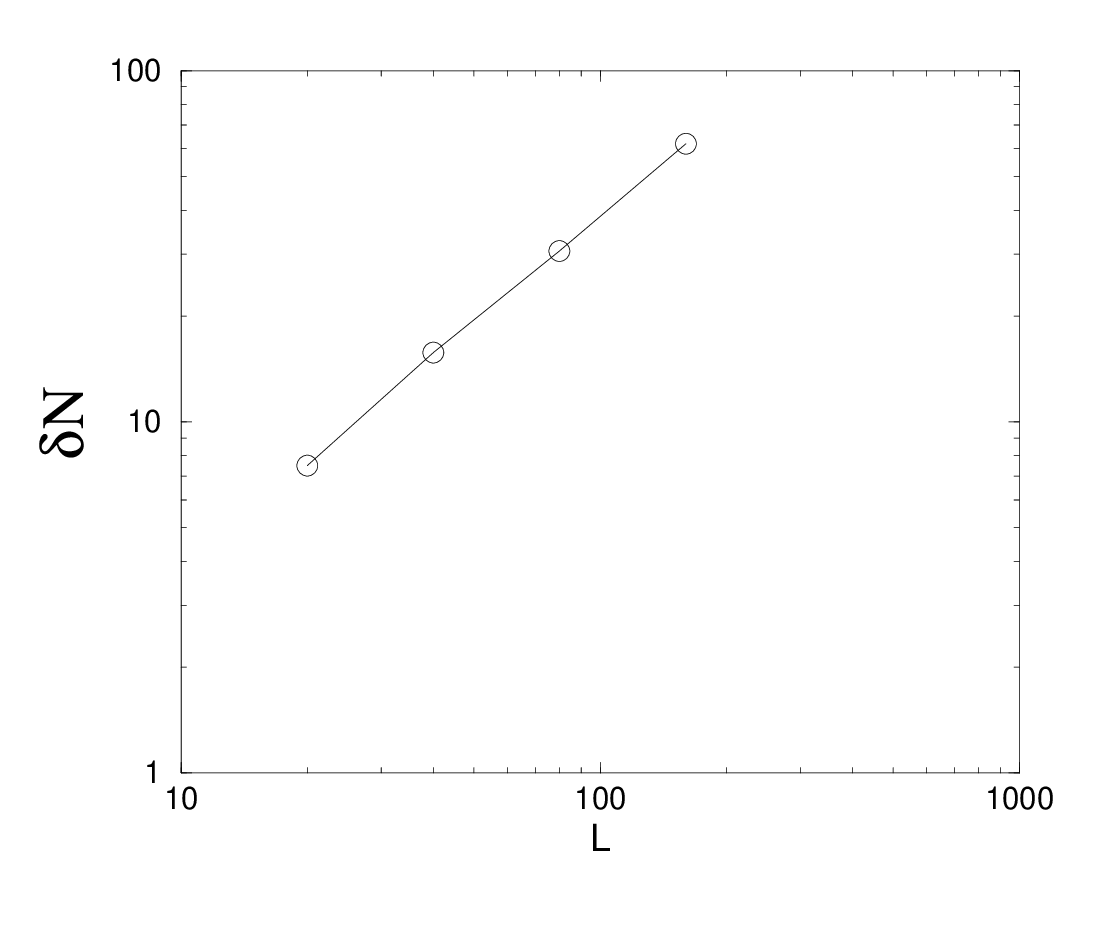}
\caption
{The dependence of excess poles number $\delta N$  on the system size L.
$\nu=0.1$ $ f^{0.5}=9e-6$}
\label{file=Fig.2}
\end{figure}
\begin{figure}
\epsfxsize=9.0truecm
\epsfbox{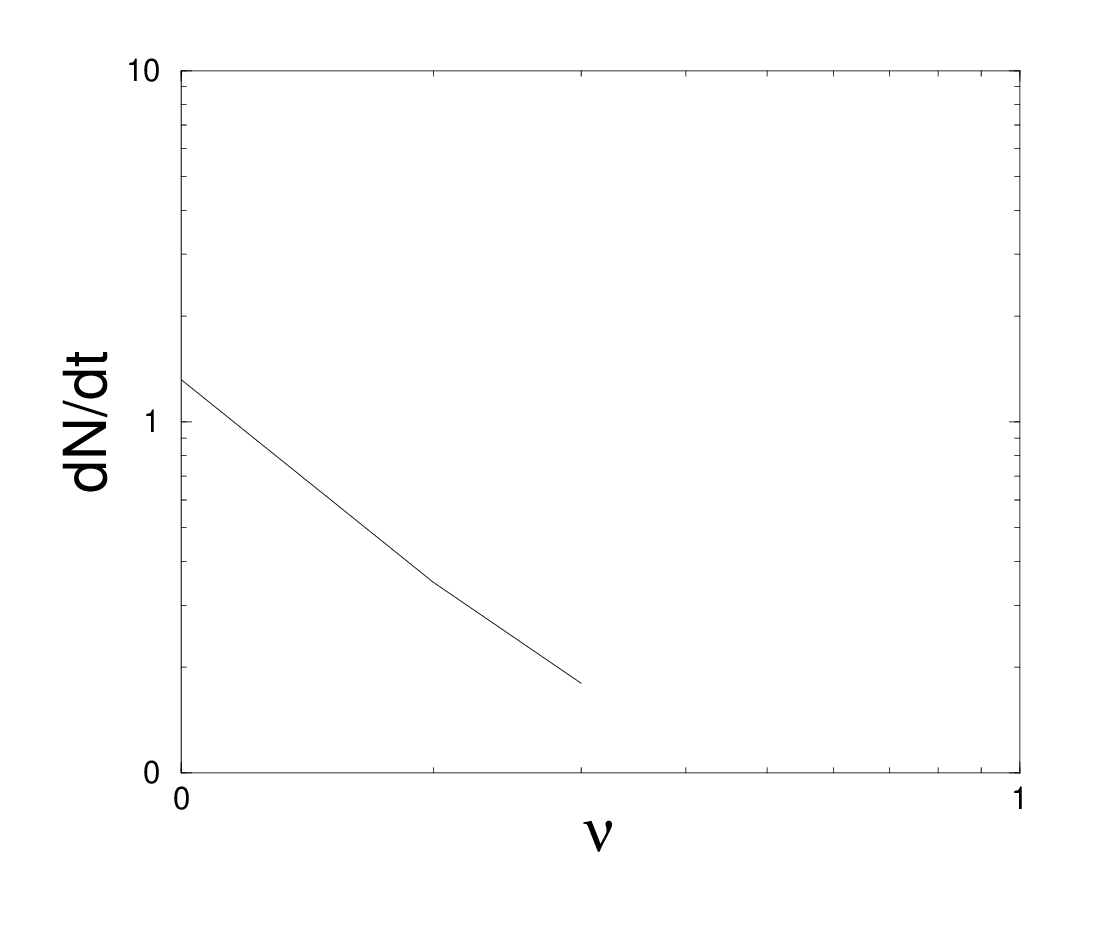}
\caption
{The dependence of poles number in time unit $dN/dt$ on the parameter $\nu$.
$L=80$ $ \nu=0.1$.}
\label{file=Fig.2}
\end{figure}
\begin{figure}
\epsfxsize=9.0truecm
\epsfbox{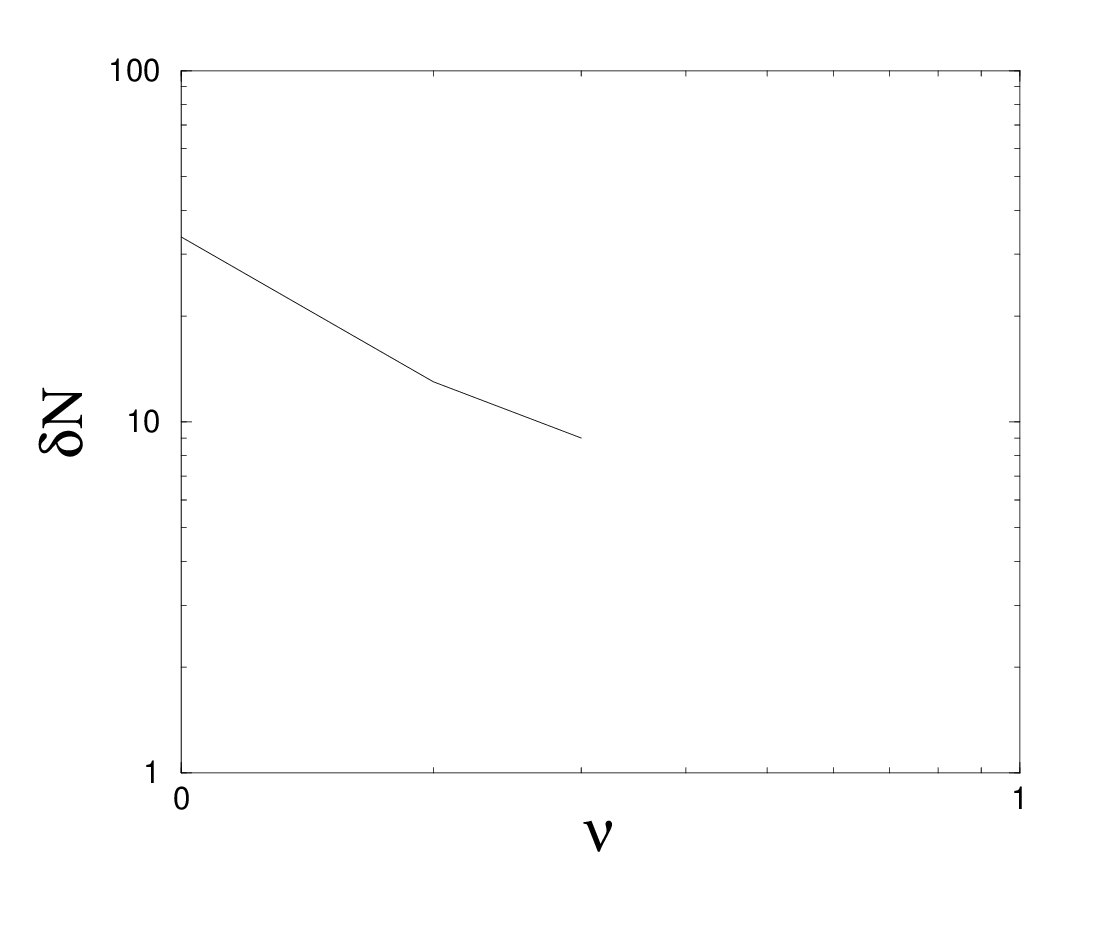}
\caption
{The dependence of excess poles number $\delta N$  on the the parameter $\nu$.
$L=80$ $ \nu=0.1$.}
\label{file=Fig.2}
\end{figure}

Saturated value of $\delta N$ is defined by next formula

\begin{equation}
\delta N \approx   N(L)/2 \approx {1 \over 4} {L \over \nu}
\end{equation}

where $N(L) \approx {1 \over 2} {L \over \nu}$ is number of poles in giant
cusp.

(iv) Regime IV We again see strong dependence of poles number  $\delta N$
on noise $f$.

\begin{equation}
\delta N \sim   f^{0.1}
\end{equation}

Because of numerical noise we can see in most of simulations only regime III
and IV. In future if we don't note something different we discuss regime III.

\item By calculation of new cusp number we can find number of poles
that appear in the system in unit of time ${ dN \over dt}$.
In regime III

\begin{equation}
{ dN \over dt} \sim   f^{0.03}
\end{equation}

Dependence on $L$ and $\nu$ define by

\begin{equation}
{ dN \over dt} \sim  L^{0.8}
\end{equation}

\begin{equation}
{ dN \over dt} \sim  {1 \over \nu^2}
\end{equation}

And in regime IV

\begin{equation}
{ dN \over dt} \sim   f^{0.1}
\end{equation}
\end{enumerate}
\subsection{Theoretical Discussion of the Effect of Noise}
\subsubsection{The Threshold of Instability to Added Noise. Transition
from regime I to regime II}
\label {regime0}
First we present the theoretical arguments that explain the sensitivity of the
giant cusp solution to the effect of added noise. This sensitivity increases
dramatically with increasing the system size $L$. To see this we use again the
relationship between the linear stability analysis and the pole dynamics.

Our additive noise introduces perturbations with all $k$-vectors. We showed
previously
that the most unstable mode is the $k=1$ component $A_1 \sin(\theta)$. Thus the
most effective noisy perturbation is $\eta_1 \sin(\theta)$ which can
potentially lead
to a growth of the most unstable mode. Whether or not this mode will grow
depends
on the amplitude of the noise. To see this clearly we return to the pole
description.
For small values of the amplitude $A_1$ we represent $A_1 \sin(\theta)$ as a
single
pole solution of the functional form $\nu e^{-y}\sin{\theta}$. The $y$
position is
determined from $y=-\log{|A_1| /\nu}$, and the $\theta$-position is
$\theta=\pi$ for
positive $A_1$ and $\theta=0$ for negative $A_1$. From the analysis of
Section III we know that
for very small $A_1$ the fate of the pole is to be pushed to infinity,
independently
of its $\theta$ position; the dynamics is symmetric in $A_1\to -A_1$ when $y$ is
large enough. On the other hand when the value of $A_1$ increases the symmetry
is broken and the $\theta$ position and
the sign of $A_1$ become very important. If $A_1>0$ there is a threshold value
of $y$ below which the
pole is attracted down. On the other hand if $A_1<0$, and $\theta=0$ the
repulsion
from the poles of the giant cusp grows with decreasing $y$. We thus
understand that
qualitatively speaking the dynamics of $A_1$ is characterized by an asymmetric
``potential" according to
\begin{eqnarray}
\dot A_1 &=& -{\partial V(A_1)\over \partial A_1}\ , \label{dvda}\\
V(A_1) &=& \lambda A_1^2 -aA_1^3+\dots \ . \label{poten}
\end{eqnarray}
>From the linear stability analysis we know that $\lambda\approx \nu/L^2$,
cf. Eq.(\ref{ya}).
We know further that the threshold for nonlinear instability is at
$A_1\approx \nu^3/L^2$,
cf. Eq(\ref{nu3L2}). This determines that value of the coefficient
$a\approx 2/3\nu^2$. The
magnitude of the ``potential" at the maximum is
\begin{equation}
V(A_{max}) \approx \nu^7/L^6 \ . \label{vmax}
\end{equation}
The effect of the noise on the development of the mode $A_1\sin{\theta}$ can be
understood from the following stochastic equation
\begin{equation}
\dot A_1 = -{\partial V(A_1)\over \partial A_1}+\eta_1(t) \ . \label{stochA}
\end{equation}
It is well known \cite{Ris} that for such dynamics the rate of escape $R$
over the ``potential" barrier
for small noise is proportional to
\begin{equation}
R\sim {\nu\over L^2} \exp^{-\nu^7/{f \over L}L^6} \ . \label{wow}
\end{equation}
The conclusion is that any arbitrarily tiny noise becomes effective when the
system size increase and when $\nu$ decreases. If we drive the system with noise
of amplitude ${f \over L}$ the system can always be sensitive to this noise when its size
exceeds a critical value $L_c$ that is determined by ${f \over L_c} \sim \nu^7/L_c^6$.
This formula defines transition from regime I (no poles) to regime II.
For $L>L_c$
the noise will introduce new poles into the system.
Even numerical noise in simulations involving large size systems may have
a macroscopic influence.

The appearance of new poles must increase the velocity of the front. The
velocity
is proportional to the mean of $(u/L)^2$. New poles distort the giant cusp by
additional smaller cusps on the wings of the giant cusp, increasing $u^2$. Upon
increasing the noise amplitude more and more smaller cusps appear in the front,
and inevitably the velocity increases. This phenomenon is discussed
quantitatively in
Section \ref{noise}.
\subsubsection{Verifing of asymmetric "potential" form}
If we know distribution of poles in the giant cusp we can find form of
"potential" and verify expressions for $\lambda$, $A_{max}$ and
${\partial V(A_1) \over \partial A_1}$.
For measurement of $A_1$ we use formula $A_1=4 \nu e^{-y}$ and
${\partial V(A_1) \over \partial A_1}= 4 \nu {dy \over dt} e^{-y}$
where ${dy \over dt}$ can be find from equation motion of poles.
Numerical measurements were made for  $ L=2 n \nu ,n-integer, n>2$.
For graphs that we use for fixed $\nu$ and variable $L$;$\nu=0.005$
$L:[1,150]$. For fixed $L$ and variable $\nu$; $L=1$ and $\nu:[0.005,0.05]$.
We can find $A_{max}$ as zero-point of ${\partial V(A_1) \over \partial A_1}$.
Obtain results are next. For ${ A_{max} L^2 \over \nu^3}$ as function of $L$
(Fig 10)
\begin{figure}
\epsfxsize=9.0truecm
\epsfbox{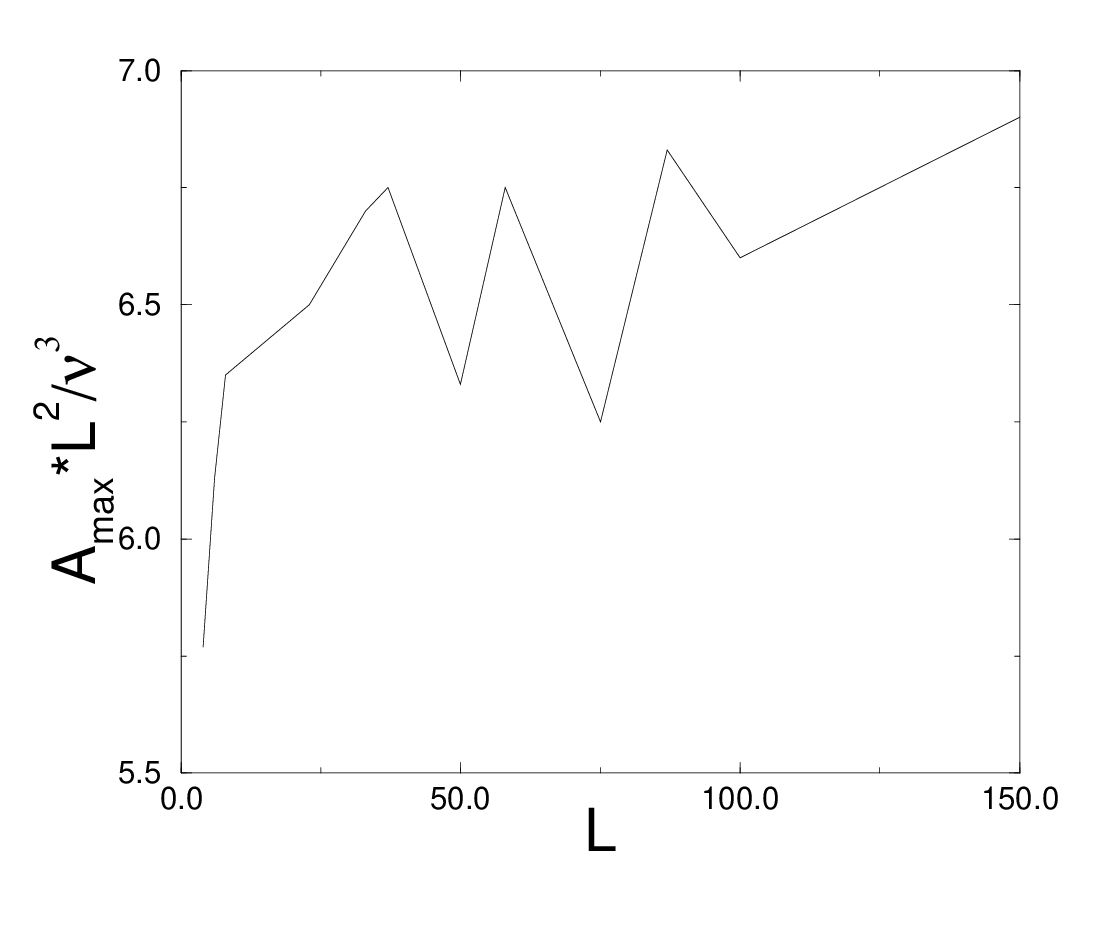}
\caption
{Dependence of the  normalized amplitude $A_{max}L^2/\nu^3$
on the system size $L$.}
\label{file=Fig.1}
\end{figure}

and $\nu$ (Fig 11)
\begin{figure}
\epsfxsize=9.0truecm
\epsfbox{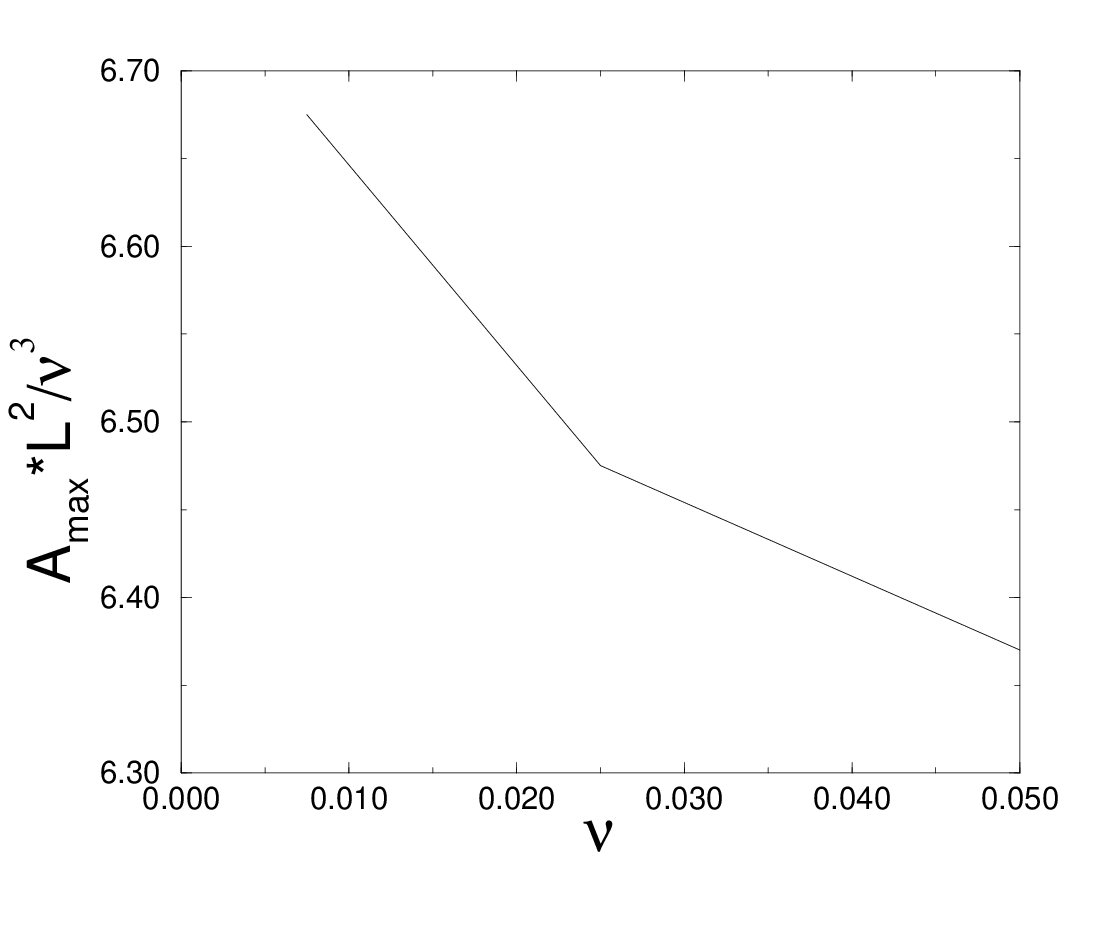}
\caption
{Dependence of the  normalized amplitude $A_{max}L^2/\nu^3$
on the parameter $\nu$.}
\label{file=Fig.1}
\end{figure}

we can see that ${ A_{max} L^2 \over \nu^3}$ is
almost constant. Dependence ${A_{max} \over A_{N(L)}}$($A_{N(L)}$ is defined by
position of the upper pole.)as function $L$
(Fig 12)
\begin{figure}
\epsfxsize=9.0truecm
\epsfbox{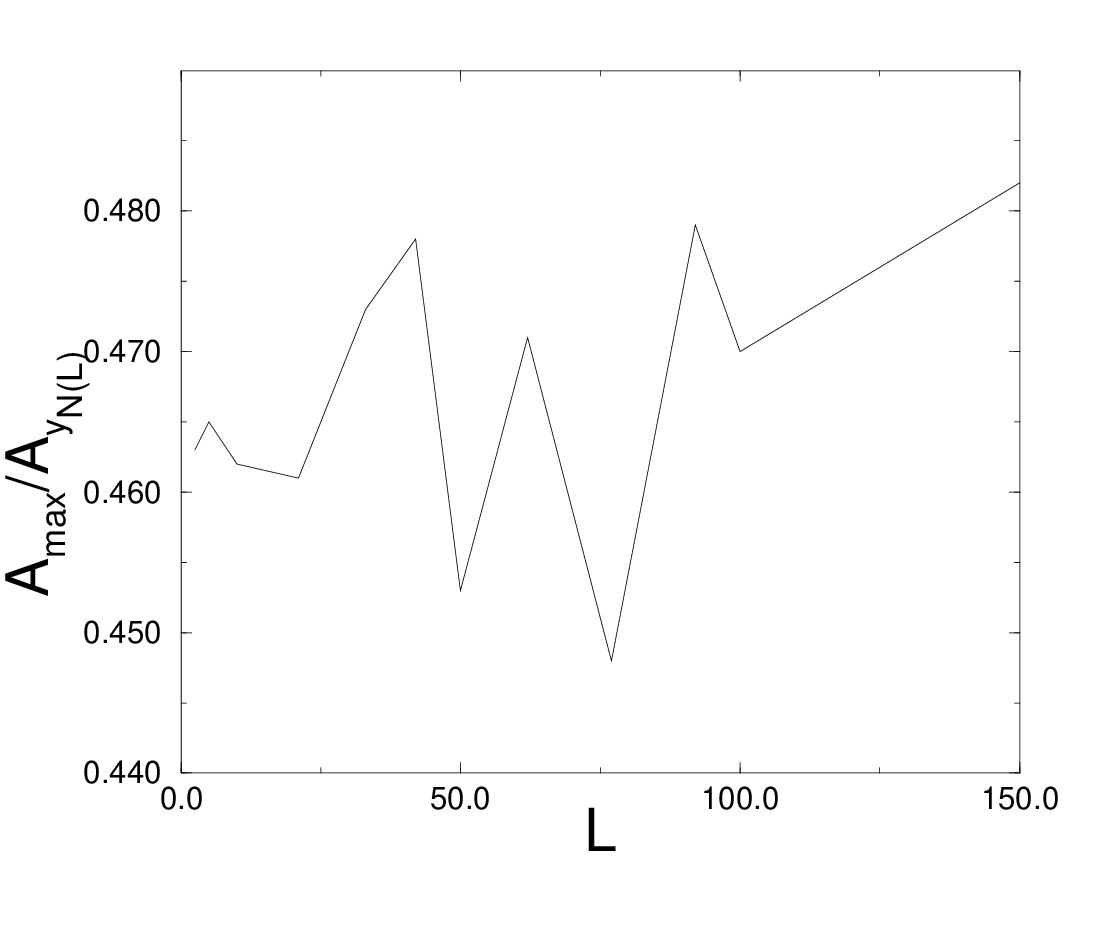}
\caption
{Relationship between amplitude defined by the minimum of potential $A_{max}$
and
amplitude defined by position of upper pole $A_{N(L)}$ as function of the
system size $L$.}
\label{file=Fig.1}
\end{figure}

and $\nu$ (Fig 13)
\begin{figure}
\epsfxsize=9.0truecm
\epsfbox{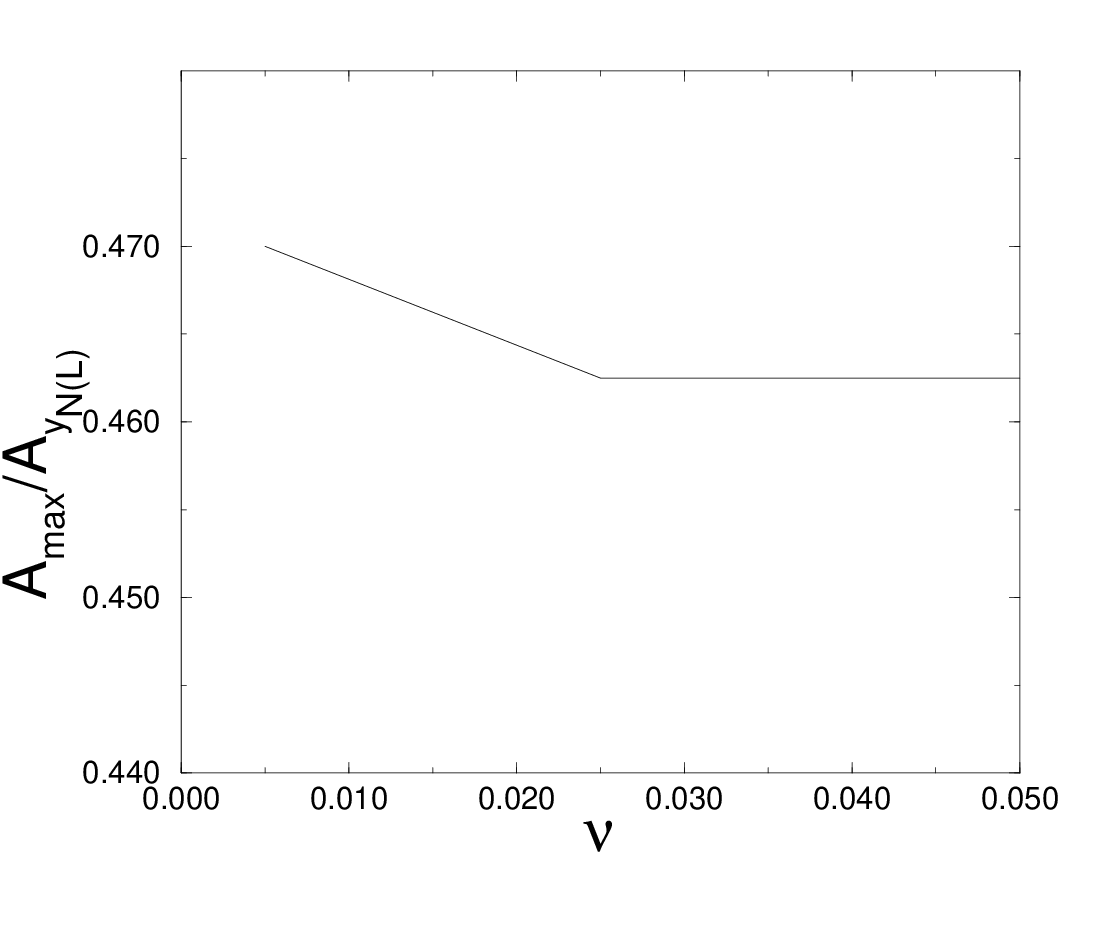}
\caption
{Relationship between amplituda defined by the minimum of potential $A_{max}$
and
amplituda defined by position of upper pole $A_{N(L)}$ as function of the
the parameter $\nu$.}
\label{file=Fig.1}
\end{figure}

is almost $Const \approx 0.5$. From graphs
Fig 14 (for $L$)
\begin{figure}
\epsfxsize=9.0truecm
\epsfbox{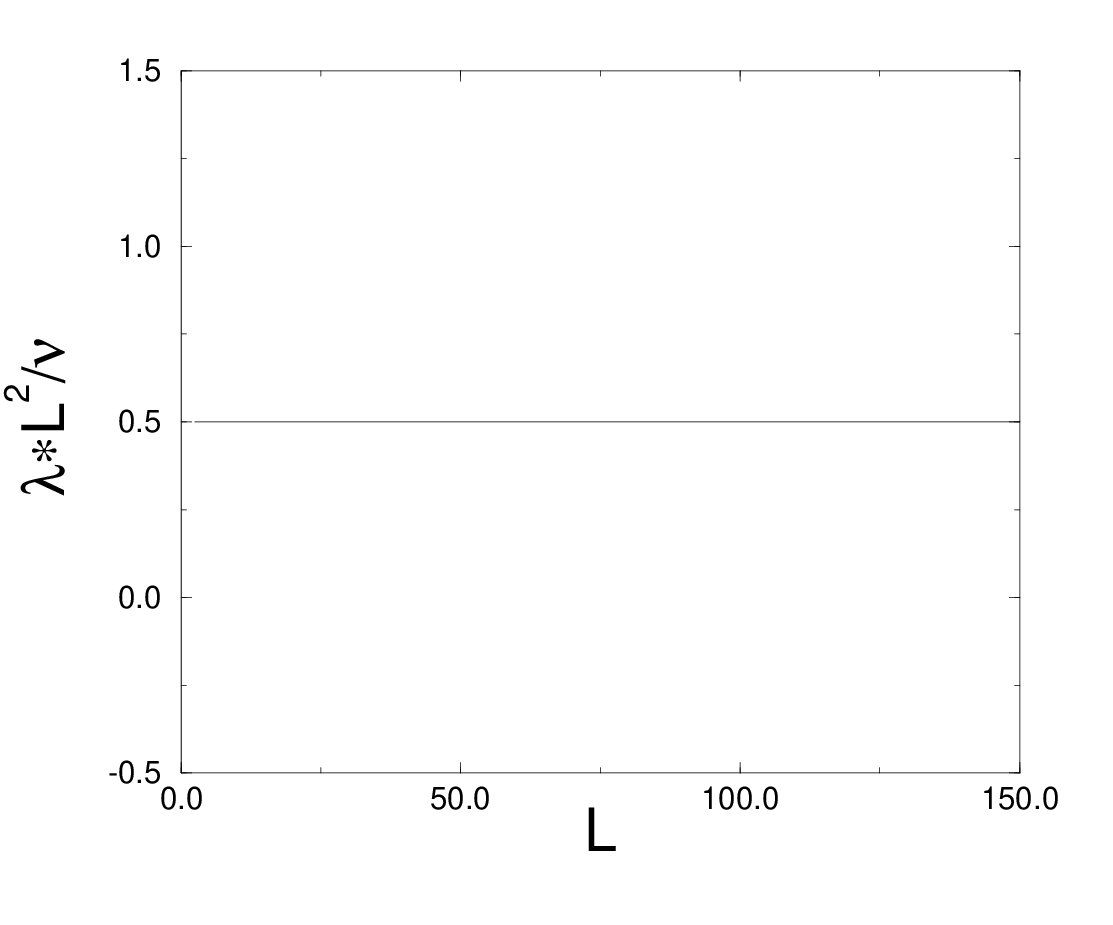}
\caption
{Dependence of the  normalized parameter $\lambda L^2/\nu$
on the system size $L$.}
\label{file=Fig.1}
\end{figure}

and Fig 15
\begin{figure}
\epsfxsize=9.0truecm
\epsfbox{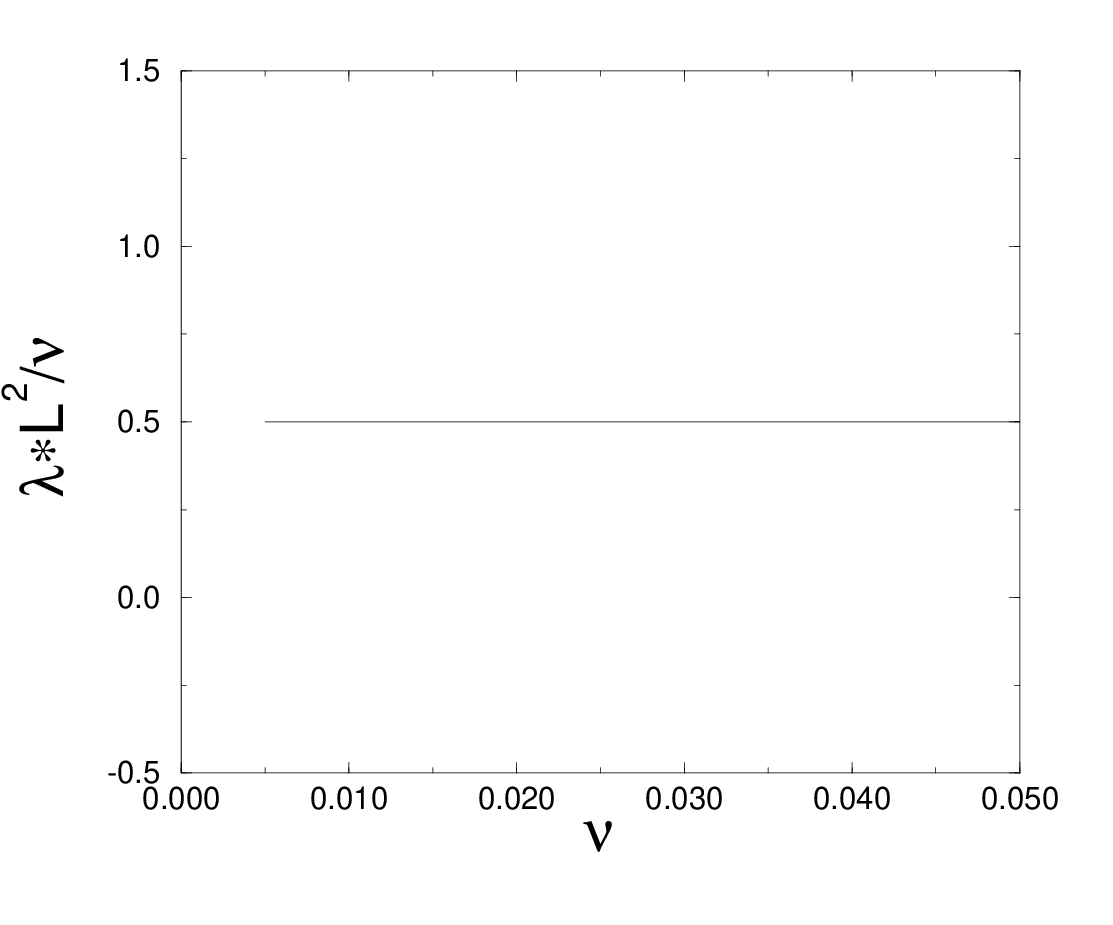}
\caption
{Dependence of the  normalized parameter $\lambda L^2/\nu$
on the parameter $\nu$.}

\label{file=Fig.1}
\end{figure}

(for $\nu$) demonstrate us that ${\lambda L^2
\over \nu}$ is almost constant.
\begin{figure}
\epsfxsize=9.0truecm
\epsfbox{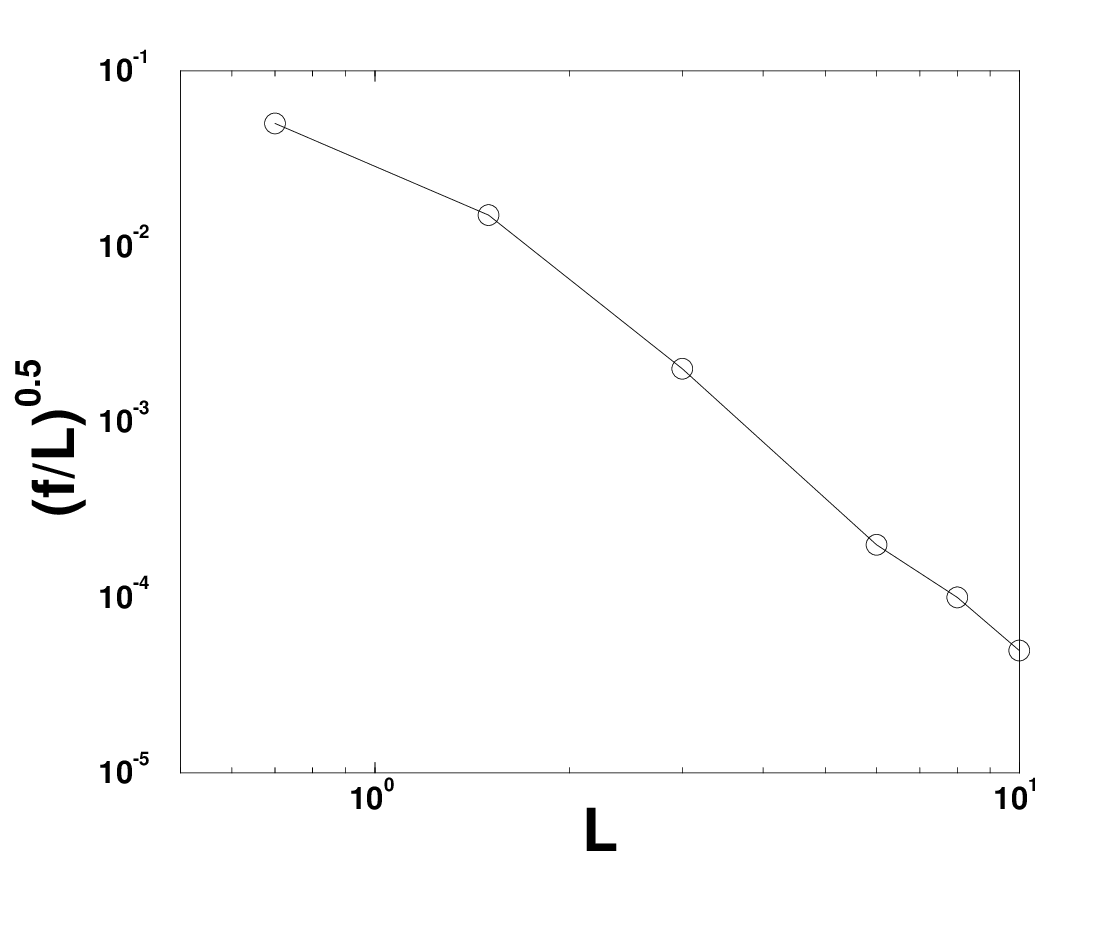}
\caption
{The first odd eigenfunction obtained from traditional stability analysis.}
\label{file=Fig.2}
\end{figure}

Fig 16 give us full dependence of ${f \over L}$
on $L$.
These results are in good agreement with the theory.

\subsubsection{The Noisy Steady State and its Collapse with
Large Noise and System Size}
In this subsection we discuss the response of the giant cusp solution to
noise levels that are able to introduce a large number of excess poles in
addition
to those existing in the giant cusp. We will denote the excess number of poles
by $\delta N$. The first question that we address is how difficult is it to
insert yet an additional pole when there is already a given excess $\delta
N$. To
this aim we estimate the effective potential $V_{\delta N}(A_1)$ which is
similar to
(\ref{poten}) but is taking into account the existence of an excess number
of poles.
A basic approximation that we employ is that the fundamental form of the
giant cusp solution is not seriously modified by the existence of an excess
number
of poles. Of course this approximation breaks down quantitatively already
with one excess pole.
Qualitatively however it holds well until the excess number of
poles
is of the order of the original number $N(L)$ of the giant cusp solution.
Another approximation is that the rest of the linear modes play no role in this case.
At this
point we limit the discussion therefore to the situation $\delta N\ll N(L)$
(regime II).

To estimate the parameter $\lambda$ in the effective potential we consider the
dynamics of one pole whose $y$ position $y_a$ is far above $y_{max}$. According
to Eq.(\ref{ya}) the dynamics reads
\begin{equation}
{dy_a\over dt}\approx {2\nu (N(L)+\delta N)\over L^2} -{1\over L}
\end{equation}
Since the $N(L)$ term cancels against the $L^{-1}$ term (cf. Sec. II A),
we remain
with a repulsive term that in the effective potential translates to
\begin{equation}
\lambda={\nu\delta N\over L^2} \ . \label{lambda2}
\end{equation}
Next we estimate the value of the potential at the break-even point between
attraction
and repulsion. In the last subsection we saw that a foreign pole has to be
inserted below $y_{max}$
in order to be attracted towards the real axis. Now we need to push the new pole
below the position of the existing  pole whose index is $N(L)-\delta N$. This
position is estimated as in Sec III C by employing the TFH distribution
function (\ref{dist}).
We find
\begin{equation}
y_{\delta N}\approx 2\ln{\Big[{4L\over \pi^2\nu\delta N}\Big]} \ . \label{ydelN}
\end{equation}
As before, this implies a threshold value of the amplitude of single pole
solution
$A_{max}\sin{\theta}$ which is obtained from equating $A_{max}=\nu
e^{-y_{\delta N}}$.
We thus find in the present case $A_{max}\sim \nu^3(\delta N)^2/L^2$. Using
again
a cubic representation for the effective potential we find $a=2/(3\nu^2\delta
N)$ and
\begin{equation}
V(A_{max}) = {1\over 3}{\nu^7(\delta N)^5\over L^6}\ . \label{max2}
\end{equation}
Repeating the calculation of the escape rate over the potential barrier we find
in the present case
\begin{equation}
R\sim {\nu\delta N\over L^2} \exp^{-\nu^7(\delta N)^5/{f \over L}L^6} \ . \label{wow2}
\end{equation}

For a given noise amplitude ${f \over L}$ there is always a value of $L$ and $\nu$ for
which
the escape rate is of $O(1)$ as long as $\delta N$ is not too large. When
$\delta N$
increases the escape rate decreases, and eventually no additional poles can
creep
into the system. The typical number $\delta N$ for fixed values of the
parameters is
estimated from equating the argument in the exponent to unity:
\begin{equation}
\delta N\approx \left({f \over L}L^6/\nu^7\right)^{1/5} \ . \label{deltaN}
\end{equation}

We can see that $\delta N$ depend on noise $f$ very seriously.
It is not the case in regime III. Let us find conditions of transition
from regime II to regime III, where we see saturation of $\delta N$
with respect to noise $f$.

(i)  We use for the amplitude of pole solution that really
equal to ${2\nu \sin \theta \over \cosh(y_{\delta N}) - \cos \theta }$
expression $A_{max}=4 \nu  e^{-y_{
\delta N}}$ but it is right only for big number $y_{\delta N}$. For
$y_{\delta N} < 1 $ better approximation is $A_{max}=
{4 \nu \over y_{\delta N}^2}$.
From equation (\ref{ydelN}) we can find that the boundary value
$y_{\delta N}=1$ correspond to $\delta N \approx N(L)/2$

(ii) We use expression $y_{\delta N} \approx
2\ln{\Big[{4L\over \pi^2\nu\delta N}\Big]}$ but for big value of
$\delta N$ better approximation that can be find the same way is
$y_{\delta N} \approx {\pi^2 \nu \over 2 L}(N(L)-\delta N)
\ln {\Big[{ 8eL \over \pi^2 \nu (N(L)-\delta N)}\Big]}$.
These expressions give us  nearly equal result for $\delta N \approx N(L)/2$.

From (i) and (ii) we can make next conclusions

(a) Transition from regime II to regime III happens for nearly
$\delta N \approx N(L)/2$

(b) Using new expression in (i) and (ii) for amplitude $A_{max}$ and
$y_{\delta N}$ we can find for noise ${f \over L}$ in regime III:

\begin{equation}
{f \over L} \sim V(A_{max}) \sim \lambda A_{max}^2 \sim {\nu \delta N \over
L^2} ({4\nu \over y_{\delta N}^2})^2 \sim {L^2 \over \nu}  {\delta N
\over (N(L)-\delta N)^4}
\end{equation}

This expression define very slow dependence of $\delta N$ on noise ${f \over L}$
for $\delta N > N(L)/2$ that explain  noise saturation of $\delta N$
for regime III.

(c) Form of the giant cusp solution is defined by poles that are
closely to zero with respect to $y$.
For regime III $N(L)/2$ poles that have position
$y < y_{\delta N=N(L)/2}=1$ stay on these place. This result explain
why giant cusp solution is not seriously modified for regime III.

From Eq.(\ref{deltaN}) by help of boundary condition
\begin{equation}
\delta N \approx
N(L)/2 \label{polla}
\end{equation}
boundary noise $f_b$ between regime II and III can be found

\begin{equation}
f_b \sim { \nu^2 } \label{nose}
\end{equation}

The basic equation describing pole dynamics is next

\begin{equation}
{dN \over dt}={\delta N \over T} \label{tlif1}
\end{equation}

where ${dN \over dt}$ is number of poles that appear in unit of time
in our system, $\delta N$ is excess number of poles, T is mean life
time of pole
 (between appearing and merging with giant cusp). Using result of numerical
simulations for ${dN \over dt}$ and (\ref{polla}) we can find for $T$

\begin{equation}
T = {\delta N \over {dN \over dt}} \sim \nu L^{0.2} \label{tlif}
\end{equation}

So life time proportional to $\nu$ and depend on system size $L$
very weakly.

\subsection{The acceleration of the flame front due to noise}
\label{accel}
In this section we estimate the scaling exponents that characterize the velocity
of the flame front as a function of the system size. Our arguments in this
section
are even less solid than the previous ones, but nevertheless we believe that we succeed
to capture
some of the essential qualitative physics that underlies the interaction between
noise and instability and which results in the acceleration of the flame front.

To estimate the velocity of the flame front we need to write down an
equation for the
mean of $<dh/dt>$ given an arbitrary number $N$ of poles in the system.
This equation follows directly from (\ref{Eqdim}):
\begin{equation}
\left<{dh\over dt}\right>={1  \over L^2}{1 \over
2\pi}\int_{0}^{2\pi}u^2d\theta  \ .
\label{eqr0}
\end{equation}
After substitution of (\ref{upoles}) in (\ref{eqr0}) we get, using (\ref{xj})
and (\ref{yj})
\begin{equation}
\left<{dh\over dt}\right>=2\nu\sum_{k=1}^N {dy_k\over dt}+2\left( {\nu N\over
     L}-{\nu^2 N^2\over L^2}\right)  \ . \label{r0pole}
\end{equation}
The estimates of the second and third terms in this equation are
straightforward.
Writing $N=N(L)+\delta N(L)$ and remembering that $N(L) \sim L/\nu $
and $\delta N(L) \sim N(L)/2$
we find that
these terms contribute $O(1)$. The first term
will contribute
only when the current of poles is asymmetric. Since noise introduces poles
at a finite
value of $y_{min}$, whereas the rejected poles stream towards infinity
and disappear at boundary of nonlinearity defined by position of highest pole

\begin{equation}
y_{max}\approx 2\ln{\Big[{4L\over \pi^2\nu}\Big]} \ . \label{delN}
\end{equation}

, we have
an asymmetry that contributes to the velocity of the front. To estimate the first
term let us define

\begin{equation}
d(\sum {dy_k\over dt})=\sum_{l}^{l+dl} {dy_k\over dt}
\end{equation}

where $\sum_{l}^{l+dl} {dy_k\over dt}$ is sum over poles that are on the
interval $y:[l,l+dl]$.
We can write

\begin{equation}
d(\sum {dy_k\over dt})=d(\sum {dy_k\over dt})_{up}-
d(\sum {dy_k\over dt})_{down}
\end{equation}

Where $d(\sum {dy_k\over dt})_{up}$ flux of poles moving up and
$d(\sum {dy_k\over dt})_{down}$ flux of poles moving down.

For these flux we can write

\begin{equation}
d(\sum {dy_k\over dt})_{up},d(\sum {dy_k\over dt})_{down} \leq {dN \over
dt} dl
\end{equation}

So for the first term

\begin{eqnarray}
&&0 \leq \sum_{k=1}^{N} {dy_k\over dt}= \\ \nonumber &&
\int_{y_{min}}^{y_{max}}{d(\sum {dy_k\over dt})
\over dl} dl\\ \nonumber &&
=\int_{y_{min}}^{y_{max}}{d(\sum {dy_k\over dt})_{up}-
d(\sum {dy_k\over dt})_{down} \over dl} dl \\ \nonumber &&
\leq {dN \over dt}(y_{max}-
y_{min})\\ \nonumber &&
\leq {dN \over dt}y_{max}
\end{eqnarray}

 Because of slow($\ln$) dependence of $y_{max}$ on $L$ and $\nu$
 ${dN \over dt}$ term define oder of nonlinearity for first term.
 This term zero for symmetric current of poles and achieves maximum
 for maximal asymmetric current of poles. Comparison $v \sim L^{0.42}f^{0.02}$
 and ${dN \over dt} \sim L^{0.8}f^{0.03}$ confirm this calculation.

\section{summary and conclusions}

The main two messages of this paper are: (i) There is an important
interaction between
the instability of developing fronts and random noise; (ii) This
interaction and its
implications can be understood qualitatively and sometimes quantitatively using
the description in terms of complex poles.

The pole description is natural in this context firstly because it
provides an
exact (and effective) representation of the steady state without noise.
Once one succeeds
to describe also the {\em perturbations} about this steady state in terms
of poles,
one achieves a particularly transparent language for the study of the
interplay between
noise and instability. This language also allows us to describe in
qualitative and
semi-quantitative terms the inverse cascade process of increasing typical
lengths when
the system relaxes to the steady state from small, random initial conditions.

The main conceptual steps in this paper are as follows: firstly one
realizes that
the steady state solution, which is characterized by $N(L)$ poles aligned along
the imaginary axis is marginally stable against noise in a periodic array of
$L$ values. For all values of $L$ the steady state is nonlinearly unstable
against noise. The main and foremost effect of noise of a given amplitude $f$ is
to introduce
an excess number of poles $\delta N(L,f)$ into the system. The existence of this
excess number of poles is responsible for the additional wrinkling of the
flame front on top of the giant cusp, and for the observed acceleration of
the flame front.
By considering the noisy appearance of new poles we rationalize the
observed scaling laws as a function of the noise amplitude and the system size.

Theoretically we therefore concentrate on estimating $\delta N(L,f)$. The
measurements
do not test our theoretical consideration directly, but rather test the
dependence of
the velocity on $L$ and $f$. The only direct test for our theory is the critical line
shown in Fig.7. The measured exponent is in accord with our analytic
estimates.
Nevertheless we note  that some of our consideration are only qualitative. For example,
we estimated $\delta N(L,f)$ by assuming that the giant cusp solution is not
seriously perturbed. On the other hand we find a flux of poles going to
infinity due
to the introduction of poles at finite values of $y$ by the noise. The
existence of poles
spread between $y_{max}$ and infinity {\em is} a significant perturbation of the
giant cusp solution. Thus also the
comparison between the various scaling exponents measured and predicted must be done
with caution; we cannot guarantee that those cases in which our prediction
hit close
to the measurement mean that the theory is quantitative. However we
believe that
our consideration extract the essential ingredients of a correct theory.

The ``phase diagram" as a function of $L$ and $f$ in this system consists of
three regimes. In the first one, discussed in
Section \ref {regime0} , the noise is
too small
to have any effect on the giant cusp solution. In the second the noise
introduces excess
poles that serve to decorate the giant cusp with side cusps. In this regime
we find
scaling laws for the velocity as a function of $L$ and $f$ and we are reasonably
successful in understanding the scaling exponents. In the third regime the
noise is large enough to create small scale structures that are not neatly
understood
in terms of individual poles. It appears from our numerics that in this
regime the
roughening of the flame front gains a contribution from the the small scale
structure in a way that is reminiscent of {\em stable}, noise driven growth
models
like the Kardar-Parisi-Zhang model.

One of our main motivations in this research was to understand the
phenomena observed
in radial geometry with expanding flame fronts. A full analysis of this problem
cannot be presented here. We note however that many of the insights offered
above
translate immediately to that problem. Indeed, in radial geometry the flame
front
accelerates and cusps multiply and form a hierarchic structure as time
progresses.
Since the radius (and the typical scale) increase in this system all the
time, new
poles will be added to the system even by a vanishingly small noise. The
marginal
stability found above holds also in this case, and the system
will allow the introduction of excess poles as a result of noise. The
results discussed
in Ref.\cite{96KOP} can be combined with the present insights to provide a theory
of radial growth. This theory will be offered in a forthcoming publication.

Finally, the success of this approach in the case of flame propagation
raises hope
that Laplacian growth patterns may be dealt with using similar ideas. A problem
of immediate interest is Laplacian growth in channels, in which a finger
steady-state solution is known to exist. It is documented that the stability
of such a finger solution to noise decreases rapidly with increasing the channel
width. In addition, it is understood that noise brings about additional
geometric
features on top of the finger. There are enough similarities here to indicate
that a careful analysis of the analytic theory may shed as much light on that
problem as on the present one.

\noindent
{\bf Acknowledgments} This work was supported in part by the German
Israeli Foundation, the US-Israel Binational Science Foundation,
the Minerva Center for Nonlinear Physics, and the
Naftali and Anna Backenroth-Bronicki Fund for Research in Chaos and
Complexity. \\~\\


\begin{thebibliography}{10}

\bibitem{Pel}
P. Pelce, {\it Dynamics of Curved Fronts}, (Academic press, Boston (1988))

\bibitem{BS}
A.-L. Barb\'{a}si and H.E. Stanley, {\it Fractal Concepts in Surface Growth}
(Cambridge University Press, 1995).

\bibitem{Vic}
T. Viscek {\it Fractal Growth Phenomena} (World Scientific, Singapore, 1992).

\bibitem{77Siv}
G.I. Sivashinsky, Acta Astronautica {\bf 4}, 1177 (1977).

\bibitem{89GIS}
Yu.A. Gostintsev, A.G. Istratov and Yu.V. Shulenin, Combust. Expl. Shock
Waves {\bf 24},
70 (1989).

\bibitem{94FSF}
L.Filyand, G.I. Sivahshinsky and M.L. Frankel,
\newblock Physica {\bf D 72}, 110 (1994).

\bibitem{96KOP}
O. Kupervasser, Z. Olami and I. Procaccia, Phys. Rev. Lett. {\bf 76}, 146
(1996).

\bibitem{82LC}
 Y. C. Lee and  H. H Chen, Phys. Scr.,{\bf T 2},41 (1982).

\bibitem{85TFH}
O. Thual, U.Frisch and M. Henon, J. Physique, {\bf 46}, 1485 (1985).

\bibitem{89Jou}
G. Joulin, J. Phys. France, {\bf 50}, 1069 (1989).

\bibitem{90Jou} G. Joulin, Zh.Eksp. Teor. Fiz., {\bf 100}, 428 (1990).

\bibitem{84SB}
B.I. Shraiman and D. Bensimon, Phys. Rev A {\bf 30}, 2840 (1984).

\bibitem{86How}
S.D. Howison, J. Fluid. Mech. {\bf 167}, 439 (1986).

\bibitem{94DMW}
S. Ponce Dawson and M. Mineev-Weinstein, Physica D{\bf 73}, 373 (1994).

\bibitem{90GS}
S. Gutman and G. I. Sivashinsky, Physica D{\bf43}, 129 (1990).
\bibitem{coars}
See  for example in:
B. Galanti, P.L. Sulem and A.D. Gilbert, Physica D {\bf47}, 416(1991).
and in the references therein.
\bibitem{Ris}
H. Risken, {\em The Fokker -Planck Equation} (Springer, Berlin 1984),
p.124 Eq.(5.111)

\end{thebibliography}
\end{document}